\documentclass[11pt,aps,showpacs,nofootinbib,prd,aps,epsf,floats,
               amsmath,amssymb,amsfonts]{revtex4-1}
\usepackage{amsmath, amssymb}
\newcommand{\mathsym}[1]{{}}

\usepackage{graphicx}
\usepackage{amsmath}
\usepackage{amssymb}
\usepackage{color}
\usepackage{bm}
\newcommand{\be}{\begin{equation}}
\newcommand{\ee}{\end{equation}}
\newcommand{\bea}{\begin{eqnarray}}
\newcommand{\eea}{\end{eqnarray}}

\newcommand{\rem}[1]{}
\newsavebox{\PSLASH}
 \sbox{\PSLASH}{$p$\hspace{-1.8mm}/}
 
\renewcommand{\theequation}{\thesection.\arabic{equation}}
\newcounter{saveeqn}
\newcommand{\add}{\addtocounter{equation}{1}}
\newcommand{\alpheqn}{\setcounter{saveeqn}{\value{equation}}%
\setcounter{equation}{0}%
\renewcommand{\theequation}{\mbox{\thesection.\arabic{saveeqn}{\alph{equation}}}}}
\newcommand{\reseteqn}{\setcounter{equation}{\value{saveeqn}}%
\renewcommand{\theequation}{\thesection.\arabic{equation}}}

 \newsavebox{\notrightarrow}
 \sbox{\notrightarrow}{$\to$\hspace{-4mm}/}
 
 \newsavebox{\PARTIALSLASH}
 \sbox{\PARTIALSLASH}{$\partial$\hspace{-1.6mm}/}
 
 \newsavebox{\ASLASH}
 \sbox{\ASLASH}{$A$\hspace{-2.1mm}/}
 
 \newsavebox{\KSLASH}
 \sbox{\KSLASH}{$k$\hspace{-1.8mm}/}
 
 \newsavebox{\LSLASH}
 \sbox{\LSLASH}{$\ell$\hspace{-1.8mm}/}
 
 \newsavebox{\QSLASH}
 \sbox{\QSLASH}{$q$\hspace{-1.8mm}/}
 
 \newsavebox{\DSLASH}
 \sbox{\DSLASH}{$D$\hspace{-2.2mm}/}
 
 \newsavebox{\DbfSLASH}
 \sbox{\DbfSLASH}{${\mathbf D}$\hspace{-2.8mm}/}
 
 \newsavebox{\DELVECRIGHT}
 \sbox{\DELVECRIGHT}{$\stackrel{\rightarrow}{\partial}$}
 
 \newcommand{\blue}{\IfColor{\textCadetBlue}{}}
\newcommand{\black}{\IfColor{\textBlack}{}}
\newcommand{\red}{\IfColor{\textRed}{}}
\newcommand{\green}{\IfColor{\textOliveGreen}{}}
\newcommand{\lila}{\IfColor{\textRedViolet}{}}

\begin{document}
\title{Top Quark Anomalous Electromagnetic Couplings in Photon-Photon Scattering at the LHC}
\author{Sh. Fayazbakhsh}\email{shfayazbakhsh@ipm.ir}
\author{S. Taheri Monfared}\email{sara.taheri@ipm.ir}
\author{M. Mohammadi Najafabadi}\email{mojtaba@ipm.ir}
\affiliation{Institute for Research in Fundamental Sciences (IPM), School of Particles and Accelerators, P.O. Box 19395-5531, Tehran-Iran}

\begin{abstract}
The capability of the LHC to study the electromagnetic dipole moments of the top quark is discussed. In particular, the process $pp\to p\gamma\gamma p\to pt\bar{t}p$, which is supposed to be tagged by the forward/backward detectors at the LHC experiments, is used to explore the top quark electric and magnetic moments. We perform analytical calculations and then a numerical analysis on the sensitivity of the total cross section of the top quark pair production in $\gamma\gamma$ scattering at the LHC to the anomalous top quark couplings with photon. It is shown that improvements in the bounds on the electromagnetic dipole moments of the top quark can be achieved in this channel in comparison with the constraints from the former studies.
\end{abstract}
 \maketitle

\section{Introduction}\label{sec1}
The Standard Model (SM) of particle physics is a well-tested effective theory, applicable at current energies, that precisely illustrates almost all experimental results and a variety of phenomena in particle physics studies, up to now. However, from the phenomenological point of view, one may attempt to model the effects of the electroweak and strong interactions of possible non-SM heavy particles, beyond TeV scale, on the experimental observables at high energy colliders. As a widely accepted framework, the effective Lagrangian approach with the content of several higher-dimension interaction terms has attracted great attention \cite{effectivelag}. According to the dimensional analysis, such new terms include expansion coefficients which are inversely proportional to powers of $\Lambda$, the scale of beyond the SM (BSM) physics.
\par
Amongst the SM elementary particles, the top quark is the heaviest one,
available now, with a mass $m_t\simeq 173.21$ GeV \cite{topmass}
nearly in the same order of the electroweak symmetry breaking energy
scale. Therefore, any deviation from the SM results is more traceable
in the top quark interactions than the other fermions ones. So far,
top quark couplings are not completely investigated and the SM
predictions for this heavy fermion have to be tested more in the case
where large numbers of top quarks and anti-quarks are produced in the
LHC.
\par
It is note-worthy that the CP violation in the SM, which is explained with a complex phase in the CKM matrix, cannot describe the matter-antimatter asymmetry in the universe due to its small amount. This asymmetry is one of the main questions in particle physics that is not answered even in the heavy quarks decay processes. Thus, the measurement of large amounts of CP violation in the top quark events can be an evidence of BSM physics. Probing the new physics effects, some of the intrinsic properties of the top quark are studied in the context of its dipole moments such as the Magnetic Dipole Moment (MDM) coming from one-loop level perturbations and the corresponding Electric Dipole Moment (EDM), which is described as a source of CP violation \cite{cpviolationtot}. Motivated by the structure of the EDM, achieved from three-loop level perturbations in the SM, one can mutate the top-gauge field vertices through CP-symmetric and CP-asymmetric anomalous form factors in order to explore the non-SM effects in the top quark pair production processes.
\par
In the SM, the top quark EDM is so small that it can be a highly attractive probe of new physics, whilst the SM prediction for the MDM of this heavy particle is not far from the upcoming experiments \cite{Bouzas:2013jha}. Although EDM and MDM quantities have been long investigated, both theoretically and experimentally, in the case of light quarks, these intrinsic properties still require more improvement on the subject of heavy quark physics. The EDM and MDM values of $\hat{a}_A<1.75\times10^{-14}$ and $\hat{a}_V=0.013$ are predicted by the SM for top quarks \cite{Hoogeveen:1990cb,Bernreuther:2005gq}. There are numerical analyses based on an extended MSSM model, including an extra vector-like multiplet, which predict the top quark EDM close to $1.75\times10^{-3}$ \cite{Ibrahim:2010hv}. Indirect measurements, stand on experimental limits on the neutron EDM, lead to upper bounds of $5.25\times10^{1}$ for the top quark EDM \cite{CorderoCid:2007uc}. A study on sensitivity of hadron colliders to constrain the top quark dipole moments within $t\bar{t}\gamma$ production has been performed in Ref. \cite{Baur:2004uw}. For the LHC at $\sqrt{s}=14$ TeV, they have reported the limit of $\pm0.2$ ($\pm0.1$) assuming the integrated luminosity $L_{\mbox{\tiny{int}}}=300$ fb$^{-1}$ ($L_{\mbox{\tiny{int}}}=3000$ fb$^{-1}$) of data. More recent limits of $-2.0\le\hat{a}_V\le0.3$ and $-0.5\le\hat{a}_A\le1.5$, coming from the branching ratio and a CP asymmetry for $b\rightarrow s\gamma$, is available in Ref. \cite{Bouzas:2012av}. The bounds of $|\hat{a}_V|<0.05~(0.09)$ and $|\hat{a}_A|<0.20~(0.28)$ are concluded from potential future measurements of $\gamma e\rightarrow t\bar{t}$ cross section with 10\% (18\%) uncertainty \cite{Bouzas:2013jha}. As will be presented in this work, our strategy leads to much more stringent limits than the current experimental ones for both EDM and MDM of the top quark.
\par
In this paper, we will focus on the top quark pair production via the Central Exclusive diffractive Production (CEP), which is defined as the process $pp \to p\gamma\gamma p\to pt\bar{t}p$, while there is no radiation between the intact outgoing beam protons and the central system $t\bar{t}$. In CEP processes, two interacting protons do not dissociate during the collision. In the simplest case, they exchange two photons and survive into the final state with extra centrally produced particle states. These protons are called forward or intact protons. The study of such events, that are classified as the forward physics studies, is becoming popular among some of the ATLAS and CMS collaborations due to the range of exclusive measurements underway at the LHC. Moreover, such collisions experimentally provide a very clean signal and represent a promising way to search for a possible BSM signal in hadron colliders.
\par
Forward protons energies can be described by means of the fractional proton energy loss, $\xi=E_{loss}/E_{p}$. Here, $E_{loss}$ is the energy that proton loses in the interaction and $E_{p}$ is the energy of the incoming proton beam. The $\xi$ parameter represents a region, which is referred to as the forward detector acceptance region, to observe intact protons in the interval $\xi_{min}<\xi<\xi_{max}$. Here, based on the CMS and ATLAS standard running conditions, we consider three different classes of the acceptance region to be \cite{Albrow:2008pn,CERN-TOTEM-NOTE}
\begin{itemize}
\item $0.0015<\xi<0.5$ (CMS-TOTEM),
\item $0.0015<\xi<0.15$ (AFP-ATLAS),
\item $0.1<\xi<0.5$ (CMS-TOTEM).
\end{itemize}
\noindent
Higher $\xi$ is available by installing forward detectors closer to the interaction points.
It is worth mentioning that the central exclusive production enables us to probe several physics subjects ranging from the SM tests to
searches for new physics such as the anomalous interactions of the gauge bosons and new heavy resonances. Several physics capability
searches can be found in \cite{cep1,cep2,cep3,cep4,cep5,cep6,cep7,cep8,cep9,cep10,cep11,cep12,taheri}.
In \cite{cep12}, the top quark flavor-changing neutral current in the $tq\gamma$ vertex with $q=u,c$ has been examined using the
$pp\rightarrow p \gamma\gamma p \rightarrow pt\bar{q}p$ process for three acceptance regions of the forward detectors.
It is shown that the sensitivity of this channel has the potential to improve the bounds on the branching ratios of $t\rightarrow q\gamma$.
\par
The present paper is organized as follows: In Sec. II A, we will introduce an effective Lagrangian for a top quark pair production process that comprises the modified interactions. Thereafter, in Sec. II B, an analytical expression for the total cross section of a diffractive collision at the LHC is provided. To demonstrate the effect of anomalous couplings on the top quark pair production process, Sec. III is devoted to a complete numerical analysis on the coupling constant dependence of the total cross section as well as the backgrounds to the Equivalent Photon Approximation (EPA) method. The constraints on the BSM couplings values are also discussed. In Sec. IV. our concluding remarks are presented.

\section{Theoretical Analysis}\label{sec2}

\subsection{The Top Quark Effective Electromagnetic Interactions}\label{sec2-1}
To clarify our framework we start with an effective Lagrangian density, involving anomalous electromagnetic couplings of top quarks which reduces to the SM Lagrangian at low energies. It has the following form \cite{saavedra}:
\begin{eqnarray}\label{Lagrangian}
{\cal{L}}_{eff.}=-g_eQ_t\bar{\psi}_t\Gamma^{\mu}_e\psi_t A_{\mu},
\end{eqnarray}
with
\begin{eqnarray}\label{vertex}
\Gamma^{\mu}_{e}=\gamma^{\mu}+\frac{i}{2m_{t}}\bigg(\hat{a}_V+i\hat{a}_A\gamma^5\bigg)\sigma^{\mu\nu}k_{\nu}.
\end{eqnarray}
Here, $A_{\mu}$ and $Q_t$ are the photon gauge field and the top quark electric charge, respectively.
The top quark field, $\psi_t$, belongs to the fundamental representation of $SU(3)_c$ color group and $g_e$ is the electromagnetic coupling constant.
The non-SM couplings are denoted by real parameters $\hat{a}_V$ and
$\hat{a}_A$, which indicate the top quark magnetic and electric
dimensionless form factors, respectively. Hence, these quantities are
proportional to the corresponding quark anomalous dipole
moments. Here, $k_{\nu}$ defines the photon four-momentum and
$m_t\simeq 173.21$ GeV is the top quark mass. The term
$\gamma^5\sigma^{\mu\nu}$, with
$\sigma^{\mu\nu}=i[\gamma^{\mu},\gamma^{\nu}]/2$, breaks the CP
symmetry, so the coefficient $\hat{a}_A$ determines the strength of a
possible CP violation process, which might originate from new
physics. In our notation we have only considered the operators with
the mass dimension $d\leq 6$ in the effective Lagrangian.

\subsection{Theoretical Calculations}\label{sec2-2}
\begin{figure}[h]
\centerline{
\includegraphics[clip,width=0.45\textwidth]{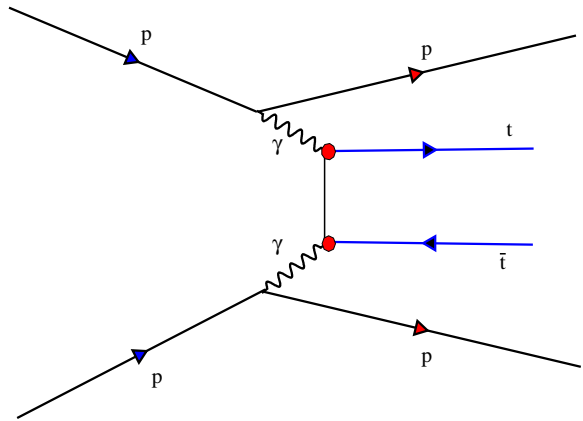}}
\caption{A schematic Feynman diagram for the two-photon exclusive top quark pair production process $pp\to p\gamma\gamma p\to pt\bar{t}p$.}
\label{F1}
\end{figure}

A representative Feynman diagram for the two-photon CEP process $pp \to p\gamma\gamma p \to pt\bar{t}p$ is illustrated in Fig. \ref{F1}. Here, two quasi-elastically incoming protons fluctuate two photons. The emitted photons can collide and produce a pair of top quarks which can be observed in the central detectors.
The scattering amplitudes for on-mass-shell photons, using the relation $\hat{s}=2m_t^2-\hat{t}-\hat{u}$ and the redefinition $R^2_{AV}=\hat{a}_A^2+\hat{a}_V^2$, read
\begin{eqnarray}\label{squaredamplitude1}
\lefteqn{\hspace{-8.4cm}|M_1|^2=\frac{8\pi^2\alpha_e^2Q_t^2}{(m_t^2-\hat{t})^2}\bigg\{m_t^4
\bigg[(R_{AV}^2+12\hat{a}_V)^2+50R_{AV}^2-72(\hat{a}_V-1)^2+56\bigg]}\nonumber\\
\lefteqn{\hspace{-4.85cm}-m_t^2\bigg[(3\hat{t}+\hat{u})\big(R_{AV}^2+6\hat{a}_V+4\big)^2+6\hat{t}\big(R_{AV}^2(17+4\hat{a}_V)+16\hat{a}_V\big)\bigg]}\nonumber\\
\lefteqn{\hspace{-4.85cm}+3\hat{t}^2\bigg[(R_{AV}^2+8\hat{a}_V)^2+34R_{AV}^2-52\hat{a}_{V}^2+32\hat{a}_V\bigg]}\nonumber\\
\lefteqn{\hspace{-4.85cm}+\hat{t}\hat{u}\bigg[3(R_{AV}^2+4\hat{a}_V)^2-12(\hat{a}_V-2)^2+16(R_{AV}^2+4)\bigg]\bigg\},}
\end{eqnarray}
\begin{eqnarray}\label{squaredamplitude2}
\lefteqn{\hspace{-8.4cm}|M_2|^2=\frac{8\pi^2\alpha_e^2Q_t^2}{(m_t^2-\hat{u})^2}\bigg\{m_t^4
\bigg[(R_{AV}^2+12\hat{a}_V)^2+50R_{AV}^2-72(\hat{a}_V-1)^2+56\bigg]}\nonumber\\
\lefteqn{\hspace{-4.85cm}-m_t^2\bigg[(\hat{t}+3\hat{u})\big(R_{AV}^2+6\hat{a}_V+4\big)^2+6\hat{u}\big(R_{AV}^2(17+4\hat{a}_V)+16\hat{a}_V\big)\bigg]}\nonumber\\
\lefteqn{\hspace{-4.85cm}+3\hat{u}^2\bigg[(R_{AV}^2+8\hat{a}_V)^2+34R_{AV}^2-52\hat{a}_{V}^2+32\hat{a}_V\bigg]}\nonumber\\
\lefteqn{\hspace{-4.85cm}+\hat{t}\hat{u}\bigg[3(R_{AV}^2+4\hat{a}_V)^2-12(\hat{a}_V-2)^2+16(R_{AV}^2+4)\bigg]\bigg\},}
\end{eqnarray}
\begin{eqnarray}\label{squaredamplitude12}
\lefteqn{\hspace{-8.4cm}M_1M^*_2+M_2M^*_1=\frac{16\pi^2\alpha_e^2Q_t^2}{(m_t^2-\hat{t})(m_t^2-\hat{u})}\bigg\{-2m_t^4
\bigg[R_{AV}^2(2R_{AV}^2+12\hat{a}_V+17)+20(\hat{a}_V+1)^2-4\bigg]}\nonumber\\
\lefteqn{\hspace{-5.4cm}+m_t^2(\hat{t}+\hat{u})\bigg[2(R_{AV}^2+5\hat{a}_V)^2+35(R_{AV}^2-\hat{a}_V)^2+(5\hat{a}_V+8)^2-80\bigg]}\nonumber\\
\lefteqn{\hspace{-5.4cm}-(\hat{t}^2+\hat{u}^2)\bigg[R_{AV}^2(4\hat{a}_V+5)+10\hat{a}_V^2+8\hat{a}_V\bigg]+
4\hat{t}\hat{u}\bigg[R_{AV}^2(R_{AV}^2-7)-5\hat{a}_V^2-16\hat{a}_V\bigg]\bigg\},}\nonumber\\
\end{eqnarray}
where, $\hat{s}=(p_1+p_2)^2=(k_1+k_2)^2$, $\hat{t}=(p_1-k_1)^2=(k_2-p_2)^2$, $\hat{u}=(k_1-p_2)^2=(p_1-k_2)^2$ and $p_i$ ($k_i$) with $i=1,2$ is the four-momentum of the final top quark (the initial photon).
\par
Before going any further with computing the total cross section we have to simplify the whole process making use of the EPA. As a proper theoretical tool, the EPA method can be applied to processes of the type $AB\rightarrow A\gamma^{*}B\rightarrow AX$, which involve virtual photons fusion. If $\gamma^{*}$ can be treated as an almost real photon, such approximation suggests that one can reduce the cross section of $AB\rightarrow AX$ to that of the subprocess $\gamma^{*}B\rightarrow X$. In a typical scattering process of a quark from a proton the corresponding amplitude is seriously sizeable in the forward direction where the exchanged photon momentum is almost zero. Consequently, in this limit, the virtual photon turns into a quasi-real photon and the EPA is justified to describe the photon spectrum in terms of its virtuality, $Q^2$, and energy, $E_\gamma$, as
\begin{eqnarray}\label{eq1}
f(E_{\gamma},Q^{2})=\frac{dN}{dE_{\gamma}dQ^2}=\frac{\alpha_e}{\pi}\frac{1}{E_{\gamma}Q^{2}}\bigg
[(1-\frac{E_{\gamma}}{E_p})(1-\frac{Q^{2}_{min}}{Q^{2}})F_{E}+\frac{E^{2}_{\gamma}}{2E_p^{2}}F_{M}\bigg].
\end{eqnarray}
In what follows, we use the relation $E_{\gamma}=E_p\xi$ and in the dipole approximation the definitions
\begin{eqnarray}\label{eq2}
&&Q^{2}_{min}=\frac{E_{\gamma}^2m_p^2}{E_p(E_p-E_{\gamma})},\nonumber\\
&&F_{E}=\frac{4m^{2}_{p}G^{2}_{E}+Q^{2}G^{2}_{M}}
{4m^{2}_{p}+Q^{2}},\;\;\;\; F_{M}=G^{2}_{M},\nonumber\\
&&G^{2}_{E}=\frac{G^{2}_{M}}{\mu^{2}_{p}}=(1+\frac{Q^{2}}{Q_{0}^{2}})^{-4},\;\;\;Q_{0}^{2}=0.71\ GeV^{2},
\end{eqnarray}
are employed \cite{budnev,Baur:2001jj,fd18,sun1} where $\alpha_e=g_e^2/4\pi$ is the fine-structure constant, $m_p$ is the proton mass, and the squared magnetic moment of the proton is taken to be a constant value $\mu^{2}_{p}=7.78$. Here, $F_{M}$ and $F_{E}$ are relative to the proton magnetic and electric form factors, respectively.
\par
In a general CEP process, the invariant mass of centrally produced particles, with energies $E_1$ and $E_2$, is obtained from $\omega\simeq 2\sqrt{E_1E_2}$. Within such events at the LHC, the invariant mass can extend to the scales which are high enough to be explored for possible new physics. In our CEP process, Fig. \ref{F1}, the produced particles are two quasi-real photons with $\omega\simeq 2\sqrt{E_{\gamma_1}E_{\gamma_2}}$. Thus, the luminosity spectrum of the emitted photons can be introduced by integrating the product of two photon spectra over the photon virtualities and energies keeping $\omega$, fixed. Evaluating the integration by changing variables from energies of two photons to $y$ and $\omega^2/4y$, where the photons virtualities remain unchanged, we arrive at the $\gamma\gamma$ luminosity spectrum
\begin{eqnarray}\label{eq3}
\frac{dL^{\gamma\gamma}}{d\omega}=\int_{y_{min}}^{y_{max}}\frac{\omega}{2y}dy\int_{Q^2_{1,min}}^{Q^{2}_{max}} dQ_1^{2}\int_{Q^{2}_{2,min}}^{Q^{2}_{max}} dQ_2^{2}\ f\big(\frac{\omega^2}{4y},Q_1^2\big)f\big(y,Q_2^2\big).
\end{eqnarray}
Virtuality of colliding photons vary between the kinematical minimum, $Q^{2}_{min}$, and a maximum, $Q^{2}_{max}\sim 1/R^{2}$, where $R$ is the proton radius \cite{fd18,Baur:2001jj}. We can conclude, from relations (\ref{eq2}), that the electric and magnetic proton form factors fall rapidly with the increase of $Q^2$. Hence, the contribution of higher virtualities, more than $Q^2_{max}=2$ GeV$^2$, to the integral (\ref{eq3}) is negligible.
\par
The total cross section that is derived by integrating $\hat{\sigma}_{\gamma\gamma\to t\bar{t}}$, the cross section of the selected subprocess, is as follows \cite{sun2}:
\begin{eqnarray}\label{cross}
\lefteqn{\hspace{-8.1cm}\sigma=\int_{\omega_{min}}^{\omega_{max}} d\omega\int_{y_{min}}^{y_{max}}\frac{\omega}{2y}dy\int_{Q^2_{1,min}}^{Q^{2}_{max}} dQ_1^{2}\int_{Q^{2}_{2,min}}^{Q^{2}_{max}} dQ_2^{2}\ f\big(\frac{\omega^2}{4y},Q_1^2\big)f\big(y,Q_2^2\big)\ \hat{\sigma}_{\gamma\gamma\to t\bar t}(Q^2_1,Q^2_2,y,\omega).}
\end{eqnarray}
In this formalism photons are supposed to be off-shell particles in order to produce top quark pairs, so $\hat{\sigma}_{\gamma\gamma\to t\bar{t}}$ in relation (\ref{cross}) also depends on photons virtualities and energies.

\section{Numerical Analysis}\label{sec3}

\subsection{The Cross Section Dependence on Top Quark Anomalous Couplings}\label{sec3-1}
\begin{figure}[h]
\centerline{
\includegraphics[clip,width=0.5\textwidth]{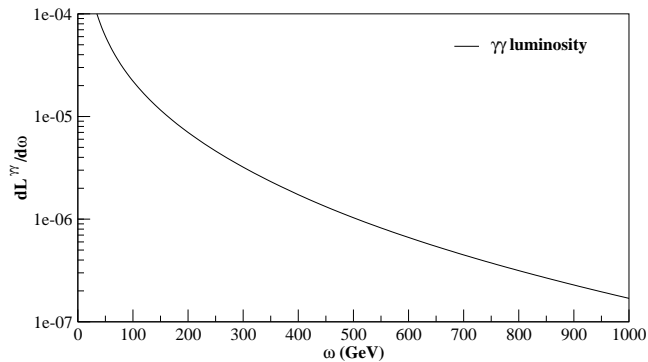}}
\caption{Photon luminosity as a function of $\omega$ at $\sqrt{s}=14$ TeV and for the acceptance region $0.0015<\xi<0.5$.}
\label{dldw}
\end{figure}

In this section, we make a numerical analysis on the anomalous electromagnetic couplings and extract the allowed regions of these parameters. At the first step, the behavior of $\gamma\gamma$ luminosity spectrum as a function of $\omega$ at $\sqrt{s}=14$ TeV and for the region $0.0015<\xi<0.5$, is shown in Fig. \ref{dldw}. The numerical calculations are performed with the integration limits
\begin{eqnarray}\label{ylimits}
&&y_{min}=Max\big[\frac{\omega^2}{4E_p\xi_{max}},E_p\xi_{min}\big],\nonumber\\
&&y_{max}=Min\big[\frac{\omega^2}{4E_p\xi_{min}},E_p\xi_{max}\big],\nonumber\\
&&\omega_{min}=Max\big[2m_t,2E_p\xi_{min}\big],\qquad \omega_{max}=\sqrt{s},
\end{eqnarray}
which are determined in terms of two boundaries of the acceptance region, $\xi_{min}$ and $\xi_{max}$. Moreover, one may use the relation $y_{max}=E_p\xi_{max}$ instead of the one in (\ref{ylimits}) and no noticeable difference appears.
\par
\begin{figure}[h]
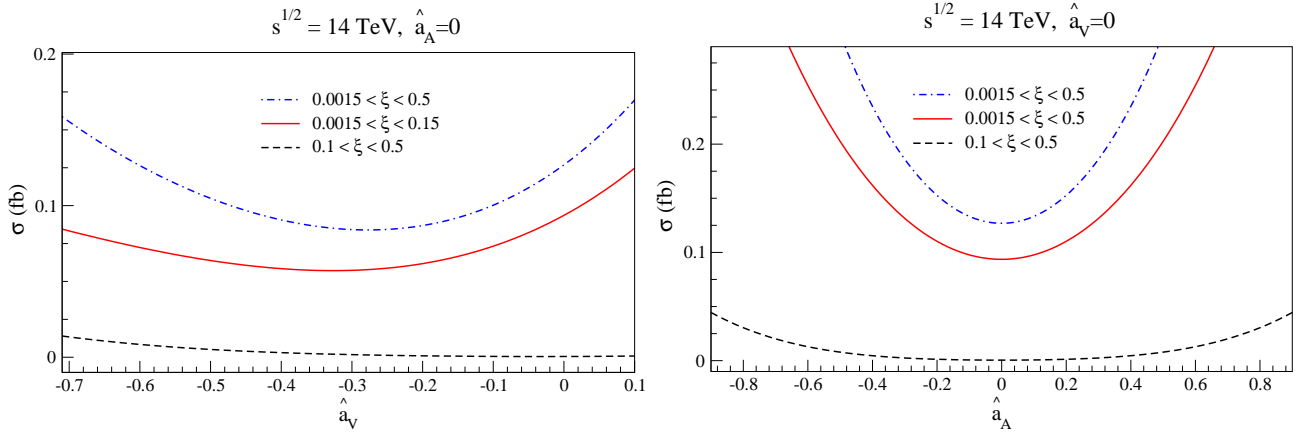

\centerline{
\includegraphics[clip,width=0.5\textwidth]{dA014-4.eps}
\includegraphics[clip,width=0.5\textwidth]{dV014-4.eps}}
\caption{The total cross section of the process $p\gamma\gamma p\to pt\bar{t}p$ as a function of the anomalous
coupling $\hat{a}_V$ at $\hat{a}_A=0$ (left panel) and $\hat{a}_A$ at $\hat{a}_V=0$ (right panel), at center of mass energy $\sqrt{s}=14$ TeV. The curves show the sensitivity for three different acceptance regions remarked on the figure.}
\label{Xsection14}
\end{figure}
\begin{figure}[h]
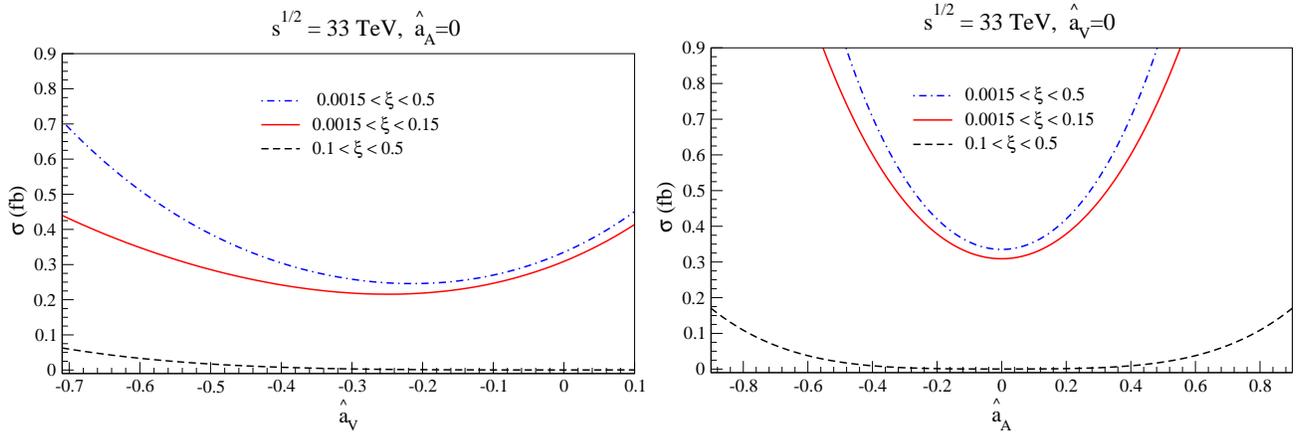

\centerline{
\includegraphics[clip,width=0.5\textwidth]{dA033-4.eps}
\includegraphics[clip,width=0.5\textwidth]{dV033-4.eps}}
\caption{The total cross section of the process $p\gamma\gamma p\to pt\bar{t}p$ as a function of the anomalous
coupling $\hat{a}_V$ at $\hat{a}_A=0$ (left panel) and $\hat{a}_A$ at $\hat{a}_V=0$ (right panel), at center of mass energy $\sqrt{s}=33$ TeV. The curves show the sensitivity for three different acceptance regions remarked on the figure.}
\label{Xsection33}
\end{figure}

In Figs. \ref{Xsection14} and \ref{Xsection33} the dependency of the total cross section to the top quark EDM and MDM is studied at the center of mass energies $\sqrt{s}=14$ and $\sqrt{s}=33$ TeV, respectively. This type of dependence is illustrated for three different acceptance detector regions by three separated curves. In each figure, left (right) panel shows the sensitivity of the cross section to the magnetic (electric) form factor, whereas the electric (magnetic) anomalous coupling is kept fixed at zero.
\par
The first acceptance region of $0.0015<\xi<0.5$ provides the most sensitive interval to the anomalous couplings. As it turns out, there is no significant difference between the results of the second region, $0.0015<\xi<0.15$, and those of the first one due to almost close cross sections. This means that the upper boundary of the acceptance region does not play indeed a major role in the total cross section value. Comparing the curves in Figs. \ref{Xsection14} and \ref{Xsection33}, we see that the sensitivity increases with increasing center of mass energy.
\par
It is analytically traceable that $\hat{\sigma}_{\gamma\gamma\to t\bar t}$ is an even function of $\hat{a}_A$ and a nonzero value of the EDM parameter always has a constructive effect on the total cross section. By contrast, in the left panel curves, there are small intervals in the vicinity of $\hat{a}_V=0$ in which the non-SM cross section is smaller than the SM one and as a result the MDM parameter has a partly destructive effect on the top quark pair production process. In each destructive interval there is a global minimum point, $\hat{a}_{V,m}$, that makes the difference $\delta\sigma(\hat{a}_V)=|\sigma_{\mbox{\tiny{SM}}}-\sigma(\hat{a}_{V})|$ maximum. In Table \ref{tab2}, the values of minima together with the corresponding cross sections, $\sigma_{\mbox{\tiny{Min}}}(\hat{a}_{V,m})$, at $\sqrt{s}=14$ TeV are presented for three acceptance regions. As an example the maximum deviation of the cross section in the second region occurs at $\hat{a}_{V,m}=-0.33$, which leads to the ratio of $\delta\sigma(\hat{a}_{V,m})/\sigma_{\mbox{\tiny{SM}}}=0.39$. In other words the possible new physics can be observed if the LHC detectors would be able to measure the cross section of the diffractive top quark pair production with a precision better than $39\%$. During a typical production process, new physics signatures may not necessarily induce an excess in the cross section rate but in some rare cases a suppression in the normal expected SM rates can also be an evidence of non-SM effects. In the case of our current work, this fact is confirmed by the destructive behavior of the anomalous MDM parameter in a small interval around $\hat{a}_V=0$.
\begin{table}[h]
\begin{tabular}{|c|c|c|c|c|c|}\hline
       $\xi$     &$\hat{a}_{V,m}$ &$\sigma_{\mbox{\tiny{SM}}}$ & $\sigma_{\mbox{\tiny{Min}}}(\hat{a}_{V,m})$ & $\delta\sigma(\hat{a}_{V,m})$ &$\delta\sigma(\hat{a}_{V,m})/\sigma_{\mbox{\tiny{SM}}}$\\\hline\hline
 0.0015-0.5 & 	 -0.28	&	0.126892 	  &		0.083981  & 0.042910	& 0.338   \\\hline
 0.0015-0.15& 	 -0.33	&	0.093597 	  &		0.057134  &	0.057134	& 0.390   \\\hline
 0.1-0.5    & 	 -0.04	&	0.000487      &		0.000469  &	0.000018	& 0.037   \\
\hline
\end{tabular}
\caption{The $\hat{a}_{V,m}$ values together with the maximum deviation of the corresponding cross sections for three forward detector acceptance regions at $\sqrt{s}=14$ TeV.}\label{tab2}
\end{table}

\subsection{Constraints on To Quark Anomalous Couplings}\label{sec3-2}
In order to obtain constraints on the top quark dipole moments, ($\hat{a}_A$,$\hat{a}_V$), a counting experiment is used. The procedure is to start with a Poisson distribution as the probability of measuring $n$ events:
\begin{equation}\label{eq:likelohood}
P(n|\,\sigma_{\mbox{\tiny{signal}}}\,\varepsilon\,L_{\mbox{\tiny{int}}}, b)=e^{-(b+\sigma_{\mbox{\tiny{signal}}}\varepsilon L_{\mbox{\tiny{int}}})}\frac{(b+\sigma_{\mbox{\tiny{signal}}}\varepsilon L_{\mbox{\tiny{int}}})^{n}}{n!}\ ,
\end{equation}
\noindent
where, $\sigma_{\mbox{\tiny{signal}}},\:\varepsilon,\:L_{\mbox{\tiny{int}}}\:$, and $b$ are the cross section of the signal in the presence of the anomalous couplings, the efficiency of the signal, the integrated luminosity, and the expected number of background events. At confidence level of $68\%$, an upper limit on the signal cross section, $\sigma_{\mbox{\tiny{signal}}}$, is calculated by integrating over the posterior probability as follows:
\begin{equation}
0.68=\frac{\int_{0}^{\sigma^{68\%}}P(n|\,\sigma_{\mbox{\tiny{signal}}}\,\varepsilon\,L_{{\mbox{\tiny{int}}}},
b)}{\int_{0}^{\infty}P(n|\,\sigma_{\mbox{\tiny{signal}}}\,\varepsilon\,L_{\mbox{\tiny{int}}},b)}\ .\label{set}
\end{equation}
\noindent
Here, $b$ denotes the number of expected events without anomalous couplings which is derived from $b=\varepsilon\times L_{\mbox{\tiny{int}}}\times Br\times \sigma_{\mbox{\tiny{bkg.}}}$. The term $\sigma_{\mbox{\tiny{bkg.}}}$ is the background cross section, i.e. $pp\to pt\bar{t}p$ in the SM framework. 
\par
To extract the expected limit on the signal cross section, one has to solve the Eq. (\ref{set}) by setting the inputs for the number of expected background events and the signal efficiency for a given integrated luminosity and branching ratio. Top quark almost always decays into a W boson and a b-quark. The decays are topologically characterized by the decay of the W boson, either leptonically, ($Br(W \rightarrow l\nu)=0.35$), or hadronically, ($Br(W\rightarrow q \bar{q})=0.65$). We consider the events that one of the top quarks decays leptonically and the other one decays either leptonically or hadronically (so called semi-leptonic and di-leptonic top quark pair events). Simultaneous hadronic decays of both top quarks is ignored to avoid large background events from the QCD production. Thus, the semi-leptonic and di-leptonic branching ratios are taken into account and the final state joint branching ratio is $Br=0.65$.
\par
The efficiency, $\varepsilon$, is the survival probability factor which is important for the predictions and it depends on the detector performance. This factor gives indeed the probability for the absence of extra inelastic interactions beside diffractive events. To obtain the efficiency, a real experimental simulation has to be done which is beyond the scope of the current paper. Although for central di-photon exchange in $\gamma\gamma$ collision $\varepsilon$ is considered to be $0.9$ \cite{cep4,Khoze:2001xm}, in the following we extend our analysis to three different efficiency values, $\varepsilon=0.1,\ 0.5,\ 0.9$.
\par
Table \ref{tab1} represents the constraints on the anomalous couplings of top quarks at $\sqrt{s}=14,33$ TeV for three different forward detector acceptance regions and integrated luminosities, $L_{\mbox{\tiny{int}}}=100,\ 300,\ 3000$ fb$^{-1}$, assuming an optimistic value for the efficiency $\varepsilon$. The first (third) acceptance region, $0.0015<\xi<0.5$ ($0.1<\xi<0.5$), is the most sensitive interval to the $\hat{a}_A$ ($\hat{a}_V$). Increasing the center of mass energy as well as the integrated luminosity provide more restricted bounds on both the anomalous couplings in all the $\xi$ values.
\begin{table}
\begin{tabular}{|c|c|c|c|c|c|}\hline
$\xi$ & $L_{\mbox{\tiny{int}}}(fb^{-1})$ & $\hat{a}_V$  & $|\hat{a}_A|$ & $\hat{a}_V$  & $|\hat{a}_A|$\\
&&$\sqrt{s}=14$ TeV&$\sqrt{s}=14$ TeV&$\sqrt{s}=33$ TeV&$\sqrt{s}=33$ TeV\\
\hline\hline
            & 100  	&-0.7542, 0.1045 & 0.2662 &-0.5237, 0.0788 & 0.1916\\

 0.0015-0.5 & 300  	&-0.6911, 0.0670 & 0.2021 &-0.4888, 0.0477 & 0.1484\\
            & 3000 	&-0.6389, 0.0233 & 0.1158 &-0.4588, 0.0168 & 0.0815\\
\hline
            & 100  	&-1.0530, 0.1234 & 0.3078 &-0.6493, 0.0788 & 0.2055\\
 0.0015-0.15& 300  	&-0.9452, 0.0780 & 0.2264 &-0.6145, 0.0542 & 0.1540\\
            & 3000 	&-0.8368, 0.0278 & 0.1342 &-0.5413, 0.0161 & 0.0899\\
\hline
            & 100  	&-0.4157, 0.3008 & 0.3467 &-0.2348, 0.2069 & 0.2237\\
 0.1-0.5    & 300  	&-0.3380, 0.2325 & 0.2773 &-0.1918, 0.1655 & 0.1796\\
            & 3000 	&-0.2128, 0.1324 & 0.1639 &-0.1218, 0.1036 & 0.1101\\
\hline
\end{tabular}
\caption{Sensitivity of the process $pp \to p\gamma\gamma p \to pt\bar{t}p$ to the top quark EDM and MDM achievable at 68\% C.L. for $\sqrt{s}=14, 33$ TeV, integrated luminosities $L_{\mbox{\tiny{int}}}=100,\ 300,\ 3000$ fb$^{-1}$, and three intervals of forward detector acceptance $\xi$. The efficiency $\varepsilon$ is taken to be $0.9$.}
\label{tab1}
\end{table}
\par
\begin{figure}[h]
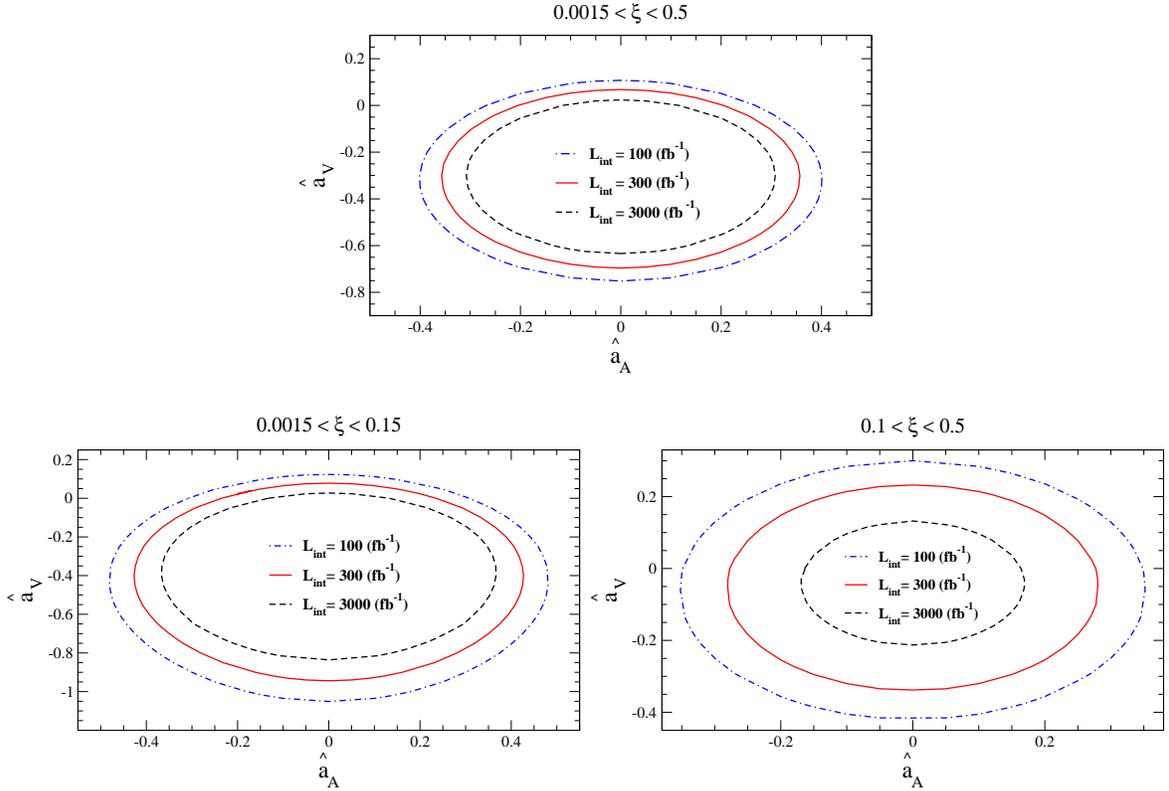

\centerline{
\includegraphics[clip,width=0.45\textwidth]{FirstregionS1s14-3.eps}}
\par
\vspace{0.5cm}
\centerline{
\includegraphics[clip,width=0.45\textwidth]{SecondregionS1s14-3.eps}
\includegraphics[clip,width=0.45\textwidth]{ForthregionS1s14-3.eps}}
\caption{The contour diagram in $\hat{a}_V-\hat{a}_A$ plane with 68\% C.L. at $\sqrt{s}=14$ TeV, $\varepsilon=0.9$, and $L_{\mbox{\tiny{int}}}=100,\ 300,\ 3000$ fb$^{-1}$. The diagrams are plotted for three different acceptance regions.}
\label{constraint}
\end{figure}

The contour diagrams for the constraints on the anomalous couplings in the $\hat{a}_V-\hat{a}_A$ plane are plotted in Fig. \ref{constraint}, at three different integrated luminosities $L_{\mbox{\tiny{int}}}=100,\ 300,\ 3000$ fb$^{-1}$ with 68\% C.L. Each panel contains the results of a specific acceptance region and, as before mentioned, there is a minor difference between the curves of the first and the second regions in which the lower boundaries are the same.
\par
\begin{figure}[h]
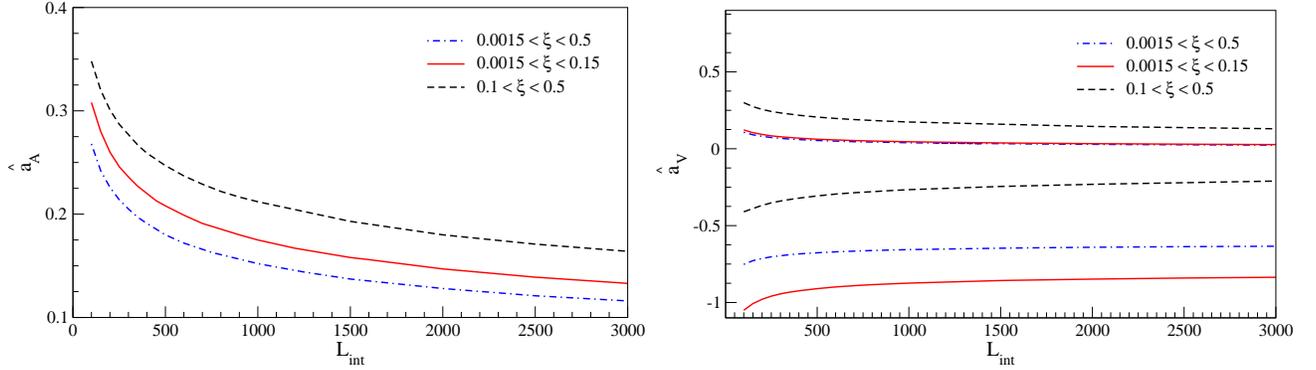

\centerline{
\includegraphics[clip,width=0.5\textwidth]{dAVersusL.eps}
\includegraphics[clip,width=0.5\textwidth]{dAAersusL.eps}}
\caption{The anomalous coupling $\hat{a}_A$ (left panel) and $\hat{a}_V$ (right panel) as a function of integrated luminosity, at $\sqrt{s}=14$ TeV and $\varepsilon=0.9$. The curves show the sensitivity for three different acceptance regions.}
\label{anomalsversL}
\end{figure}

In Fig. \ref{anomalsversL}, left panel (right panel), the behavior of the limit on the anomalous coupling $\hat{a}_A$ ($\hat{a}_V$) at $68\%$ CL as a function of integrated luminosity, at center of mass energy $\sqrt{s}=14$ TeV, is depicted. To extract the above bounds on $\hat{a}_V$ and $\hat{a}_A$, we consider the SM top quark pair production of $pp\to p\gamma\gamma p\to pt\bar{t}p$ as the background and assume the efficiency of observing signal and background events to be $\varepsilon=0.9$. The curves are presented for three different forward detector acceptance regions. As it can be seen, the $68\%$ CL sensitivity on the top quark EDM is more than its MDM to the amount of data. This is because of the stronger dependence of the signal cross section on $\hat{a}_A$ than $\hat{a}_V$. The most sensitive region to $\hat{a}_A$ ($\hat{a}_V$) parameter is the first (third) acceptance interval.
\par
To have a detailed study on the cross section behavior, we present the analytical form of our theoretical cross section in terms of the anomalous couplings,
\begin{eqnarray}\label{eq:fit}
\sigma(\hat{a}_V)&=&\alpha\hat{a}_V^4+\gamma\hat{a}_V^3+\beta\hat{a}_V^2+\rho\hat{a}_V+\eta,\nonumber\\
\sigma(\hat{a}_A)&=&\alpha^{\prime}\hat{a}_A^4+\beta^{\prime}\hat{a}_A^2+\eta.
\end{eqnarray}
The expansion coefficients, arising from the structure of the scattering amplitudes in (\ref{squaredamplitude1})-(\ref{squaredamplitude12}), are summarized in Tables \ref{tab3}. The parameter $\eta$ is the cross section at $\hat{a}_A=\hat{a}_V=0$ that represents the SM cross section magnitude.
\begin{table}[h]
\begin{tabular}{|c|c|c|c|c|c|c|c|}\hline
     $\xi$     & $\alpha$ & $\gamma$  & $\beta$ & $\rho$  & $\eta$ & $\alpha'$ & $\beta'$ \\\hline\hline
 0.0015-0.5 & 0.255 	&0.549  & 0.805  &0.341  & 0.127  & 0.255 	&0.635 \\\hline
 0.0015-0.15& 0.0953  	&0.348  & 0.539  &0.254  & 0.0936 & 0.0953  &0.412\\\hline
 0.1-0.5    & 0.0443  	&0.0182 & 0.0190 &0.0012 & 0.00049 & 0.0443  	&0.0185\\
\hline
\end{tabular}
\caption{Numerical fitted parameters for the cross section versus $\hat{a}_V$ and $\hat{a}_A$.}
\label{tab3}
\end{table}
\par
According to Eq. \ref{eq:fit} and Fig. \ref{Xsection14}, the bounds on anomalous couplings receive unequal contributions from two different features of the total cross section: the number of contributing background events, $\sigma_{\mbox{\tiny{SM}}}$, and the slope of its changes with respect to $\hat{a}_A$ and $\hat{a}_V$ . It is the latter quantity that plays the main role on the bounds sensitivity in such a way that the existence of different slopes on two sides of the vertical axes in Fig. \ref{Xsection14}, leads to different upper and lower bounds on EDM and MDM.
\par
\begin{figure}[h]
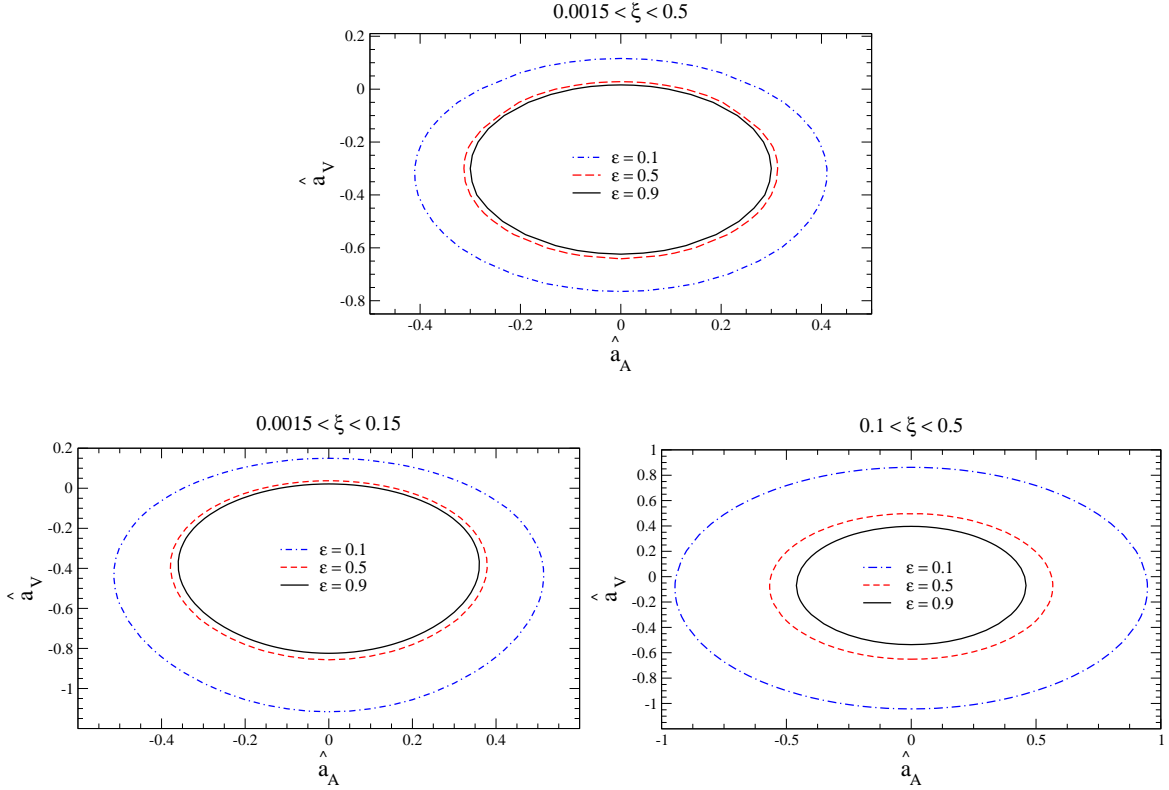

\centerline{
\includegraphics[clip,width=0.45\textwidth]{dAdVFirsteffi.eps}}
\par
\vspace{0.5cm}
\centerline{
\includegraphics[clip,width=0.45\textwidth]{dAdVSecondeffi.eps}
\includegraphics[clip,width=0.45\textwidth]{dAdVFortheffi.eps}}
\caption{The contour diagram in $\hat{a}_V-\hat{a}_A$ plane with 68\% C.L. at $\sqrt{s}=14$ TeV, $L_{\mbox{\tiny{int}}}=3000$ fb$^{-1}$, and $\varepsilon=0.1,\ 0.5,\ 0.9$. Total background has been taken 10 times more than signal. The diagrams are plotted for three different acceptance regions.}
\label{constrainteffi}
\end{figure}

To check the limits dependency upon the efficiency and to be more conservative on the background contributions, we get the limits on $\hat{a}_V$ and $\hat{a}_A$ for three efficiency values of $0.1,\ 0.5\ $, and $0.9$ under the assumption that the number of background is 10 times more than that of the signal, i.e. $b=10\times s$. The effect of the efficiency reduction and the conservative assumption for background contributions on the bounds of $\hat{a}_V$ and $\hat{a}_A$ at 68\% C.L., is illustrated in Fig. \ref{constrainteffi}. The results are presented for three forward detector acceptance regions and integrated luminosity $L_{\mbox{\tiny{int}}}=3000$ fb$^{-1}$.
\par
As it was expected, the limits get looser with respect to the previous shown results in Table \ref{tab1}, due to the smaller number of signal events and larger background contributions. Comparing the results of the optimal first detector acceptance region in Fig. \ref{constrainteffi} with the corresponding data in Fig. \ref{constraint} at $L_{\mbox{\tiny{int}}}=3000$ fb$^{-1}$, one can conclude that increasing the amount of background by a factor of 10 leads to looser bounds by a factor of around 2-3. Decreasing the efficiency from $0.9$ to $0.5$ does not significantly affect the limits while going down to $0.1$  non-negligibly loosen the bounds on the electromagnetic moments of the top quark. Even in such a pessimistic case, the limits are comparable with the ones from other studies mentioned in the first section.
\section{Concluding Remarks}\label{sec4}
The LHC is going to collide proton beams and get data for its second run. It allows to study a new energy and luminosity beyond the capabilities of the previous particle colliders. In addition to a proton-proton collider, LHC can
be considered as a photon-photon collider where not only the SM could be
examined but also several physics beyond the SM could be tested with
small backgrounds and a clean environment due to the absence of proton beam
remnants. The $\gamma\gamma$ fusion enables us to probe in particular
the electromagnetic properties of the SM particles.
\par
In this paper, we have explored the phenomenology of anomalous
$\gamma t\bar{t}$ couplings in the subprocess $\gamma\gamma\to
t\bar{t}$ in $pp$ collisions at the center of mass energies $\sqrt{s}=14,\ 33$ TeV and integrated luminosities $L_{\mbox{\tiny{int}}}=100,\ 300,\ 3000$ fb$^{-1}$. We calculate the cross section of the process $pp\to p\gamma\gamma p\to pt\bar{t}p$ for three different detector acceptance regions, $0.0015<\xi<0.5$, $0.0015<\xi<0.15$, and $0.1<\xi<0.5$. We have found more sensitivity to the top quark EDM with respect
to the MDM in this process. The detector acceptance region of $0.0015<\xi<0.5$ provides stronger bounds. The $68\%$ CL
bounds on the top quark electric and magnetic dipole moments at the
LHC with the center of mass energy of 14 TeV are found to be in the
intervals of $(-0.1158,0.1158)$ and $(-0.6389, 0.0233)$ using $L_{\mbox{\tiny{int}}}=3000$ fb$^{-1}$ of data, respectively. Finally, it should be mentioned that to obtain more realistic bounds on the anomalous $\gamma t\bar{t}$ couplings, all sources of backgrounds as well as detector effects have to be considered by the experimental collaborations. 

\section*{ACKNOWLEDGMENTS}\label{sec5}
Sh. F and S. T. M. acknowledge M. Zeinali for valuable discussions on the scattering amplitudes.



\begin{thebibliography}{99}

\bibitem{effectivelag}
  B. Henning, X. Lu and H. Murayama, arXiv:1412.1837 [hep-ph];
  \par
  J. Gonzalez-Fraile, arXiv:1411.5364 [hep-ph];
  \par
  L. Lehman Phys.\ Rev.\ D {\bf 90}, no. 12, 125023 (2014) [arXiv:1410.4193 [hep-ph]];
  \par
  C. Englert and M. Spannowsky, Phys.\ Lett.\ B {\bf 740}, 8 (2014) [arXiv:1408.5147 [hep-ph]];
  \par
  E. Masso, JHEP {\bf 1410}, 128 (2014) [arXiv:1406.6376 [hep-ph]];
  \par
  M. B. Einhorn and J. Wudka, Nucl.\ Phys.\ B {\bf 876}, 556 (2013) [arXiv:1307.0478 [hep-ph]];
  \par
  C. Degrande, N. Greiner, W. Kilian, O. Mattelaer, H. Mebane, T. Stelzer, S. Willenbrock and C. Zhang, Annals Phys.\ {\bf 335}, 21 (2013) [arXiv:1205.4231 [hep-ph]];
  \par
  J. F. Kamenik, J. Shu and J. Zupan, Eur.\ Phys.\ J.\ C {\bf 72}, 2102 (2012) [arXiv:1107.5257 [hep-ph]];
  \par
  C. Zhang and S. Willenbrock, Phys.\ Rev.\ D {\bf 83}, 034006 (2011) [arXiv:1008.3869 [hep-ph]];
  \par
  B. Grzadkowski, M. Iskrzynski, M. Misiak and J. Rosiek, JHEP {\bf 1010}, 085 (2010) [arXiv:1008.4884 [hep-ph]];
  \par
  W. Buchmuller and D. Wyler, Nucl.\ Phys.\ B {\bf 268}, 621 (1986).

\bibitem{topmass}
  K.~A.~Olive {\it et al.}  [Particle Data Group Collaboration],
  Chin.\ Phys.\ C {\bf 38}, 090001 (2014).

\bibitem{cpviolationtot}
  W. Geng, FERMILAB-THESIS-2012-46;
  \par
  S. Lee, FERMILAB-THESIS-2011-07;
  \par
  Y.~Sumino, hep-ph/0007326.
  \par
  D. Atwood, S. Bar-Shalom, G. Eilam and A. Soni, Phys.\ Rept.\  {\bf 347}, 1 (2001) [hep-ph/0006032];
  \par
  A. Brandenburg and J. P. Ma, 
   Phys.\ Lett.\ B {\bf 298}, 211 (1993);

\bibitem{Bouzas:2013jha}
  A.~O.~Bouzas and F.~Larios,
  Phys.\ Rev.\ D {\bf 88}, no. 9, 094007 (2013)
  [arXiv:1308.5634 [hep-ph]].

\bibitem{Hoogeveen:1990cb}
  F.~Hoogeveen,
  Nucl.\ Phys.\ B {\bf 341}, 322 (1990).

\bibitem{Bernreuther:2005gq}
  W.~Bernreuther, R.~Bonciani, T.~Gehrmann, R.~Heinesch, T.~Leineweber, P.~Mastrolia and E.~Remiddi,
  Phys.\ Rev.\ Lett.\  {\bf 95}, 261802 (2005)
  [hep-ph/0509341].

\bibitem{Ibrahim:2010hv}
  T.~Ibrahim and P.~Nath,
  Phys.\ Rev.\ D {\bf 82}, 055001 (2010)
  [arXiv:1007.0432 [hep-ph]].

\bibitem{CorderoCid:2007uc}
  A.~Cordero-Cid, J.~M.~Hernandez, G.~Tavares-Velasco and J.~J.~Toscano,
  J.\ Phys.\ G {\bf 35}, 025004 (2008)
  [arXiv:0712.0154 [hep-ph]].

\bibitem{Baur:2004uw}
  U.~Baur, A.~Juste, L.~H.~Orr and D.~Rainwater,
  Phys.\ Rev.\ D {\bf 71}, 054013 (2005)
  [hep-ph/0412021].

\bibitem{Bouzas:2012av}
  A.~O.~Bouzas and F.~Larios,
  Phys.\ Rev.\ D {\bf 87}, no. 7, 074015 (2013)
  [arXiv:1212.6575 [hep-ph]].

\bibitem{Albrow:2008pn}
  M.~G.~Albrow {\it et al.}  [FP420 R and D Collaboration],
  JINST {\bf 4}, T10001 (2009)
  [arXiv:0806.0302 [hep-ex]].

\bibitem{CERN-TOTEM-NOTE}
V. Avati and K. Osterberg. 2005. Report No. CERN-TOTEM-NOTE--002, (2006).

\bibitem{cep1}
The CMS and TOTEM Collaborations, CERN/LHCC 2006-039/G-124.

\bibitem{cep2}
 A.~De Roeck, V.~A.~Khoze, A.~D.~Martin, R.~Orava and M.~G.~Ryskin,
  Eur.\ Phys.\ J.\ C {\bf 25}, 391 (2002)
  [hep-ph/0207042].

\bibitem{cep3}
D.~G.~d'Enterria,
  arXiv:0708.0551 [hep-ex].

\bibitem{cep4}
O.~Kepka and C.~Royon,
  Phys.\ Rev.\ D {\bf 78}, 073005 (2008)
  [arXiv:0808.0322 [hep-ph]].

\bibitem{cep5}
 S.~C.~Inan,
  Phys.\ Rev.\ D {\bf 81}, 115002 (2010)
  [arXiv:1005.3432 [hep-ph]].

\bibitem{cep6}
E.~Chapon, C.~Royon and O.~Kepka,
  Phys.\ Rev.\ D {\bf 81}, 074003 (2010)
  [arXiv:0912.5161 [hep-ph]].

\bibitem{cep7}
 S.~Atag and A.~A.~Billur,
  JHEP {\bf 1011}, 060 (2010)
  [arXiv:1005.2841 [hep-ph]].

\bibitem{cep8}
 S.~C.~Inan and A.~A.~Billur,
  Phys.\ Rev.\ D {\bf 84}, 095002 (2011).

\bibitem{cep9}
  A. A. Billur, Europhys.\ Lett.\  {\bf 101}, 21001 (2013).

\bibitem{cep10}
L.~N.~Epele, H.~Fanchiotti, C.~A.~G.~Canal, V.~A.~Mitsou and V.~Vento,
  Eur.\ Phys.\ J.\ Plus {\bf 127}, 60 (2012)
  [arXiv:1205.6120 [hep-ph]].

\bibitem{cep11}
 M.~Kosal and S.~C.~Inan,
  Adv.\ High Energy Phys.\  {\bf 2014}, 315826 (2014)
  [arXiv:1403.2760 [hep-ph]].

\bibitem{cep12}
 S.~C.~Inan,
  arXiv:1410.3609 [hep-ph].

 \bibitem{taheri}  S. Taheri Monfared and S. Fayazbakhsh, Acta Phys.\ Polon.\ Supp.\  {\bf 7}, no. 3, 579 (2014).

\bibitem{saavedra}
 J.~A.~Aguilar-Saavedra,
  Nucl.\ Phys.\ B {\bf 812}, 181 (2009)
  [arXiv:0811.3842 [hep-ph]].

\bibitem{budnev}
  V. M. Budnev, I. F. Ginzburg, G. V. Meledin and V. G. Serbo, Phys.\ Rept.\  {\bf 15}, 181 (1975).

\bibitem{fd18}
  K. Piotrzkowski, Phys. Rev. D {\bf 63}, 071502 (2001) [arXiv:hep-ex/0009065].


\bibitem{sun1}
  I. Sahin and M. Koksal, JHEP {\bf 1103}, 100 (2011) [arXiv:1010.3434 [hep-ph]].

\bibitem{Baur:2001jj}
  G.~Baur, K.~Hencken, D.~Trautmann, S.~Sadovsky and Y.~Kharlov,
  Phys.\ Rept.\  {\bf 364}, 359 (2002)
  [hep-ph/0112211].

\bibitem{sun2}
  H. Sun, Eur.\ Phys.\ J.\ C {\bf 74}, no. 8, 2977 (2014) [arXiv:1406.3897 [hep-ph]].

\bibitem{Khoze:2001xm}
  V.~A.~Khoze, A.~D.~Martin and M.~G.~Ryskin,
  Eur.\ Phys.\ J.\ C {\bf 23}, 311 (2002)
  [hep-ph/0111078].





\end{thebibliography}
\end{document}